\documentstyle[12pt,makeidx,epsfig,rotating]{article}
\newcommand{\bp}{\begin{picture}}
\newcommand{\ep}{\end{picture}}
\include{def}
\def\titlefoot#1{\stepcounter{footnote}\footnotetext{#1}}
\newcommand{\Aname}[2]{ #1$^{\ref{#2}}$}
\begin{document}
\begin{titlepage}
\par 
\begin{center}
{\large EUROPEAN ORGANIZATION FOR NUCLEAR RESEARCH \\
\bigskip
\hfill 
\hfill \normalsize CERN-EP/98-56 (revised)} \\
{\hfill  \normalsize April 9, 1998} \\
{\hfill  \normalsize revised June 22, 1998} \\
%
%
\vfill 
{\Large \bf Pseudo-Dirac neutrinos as a potential complete solution to the
  neutrino oscillation puzzle}
\vspace*{0.3cm}
\vfill 
\vspace*{0.4cm}
\normalsize 
\Aname{A. Geiser}{CERN}
\vfill 
\begin{abstract}
\indent 
\textwidth 8.0cm 
\hoffset 5.0cm 
A solution for the neutrino mass and mixing pattern is proposed which is 
compatible with all available experimental data on neutrino oscillations. 
This solution involves Majorana neutrinos of the pseudo-Dirac type, i.e.
$m_{\rm Majorana} \ll m_{\rm Dirac}$. 
The solar and atmospheric neutrino observations
are mainly explained as $\nu_e - \nu_e^S$ and $\nu_\mu - \nu_\mu^S$ 
oscillations,
where S indicates the sterile (``righthanded'') partner of each neutrino 
generation, while the LSND result is interpreted in terms of 
standard $\nu_\mu - \nu_e$ oscillations. The resulting constraints on  
$\nu_\mu - \nu_\tau$ and $\nu_\tau - \nu_\tau^S$ oscillations are also
discussed. This solution leaves room for
a hierarchical mass and mixing scheme with a $\nu_\tau$ mass in the few eV
range, as favoured by some dark matter scenarios. 
The apparent conflict with standard Big Bang nucleosynthesis is addressed
and the implications for current and future experiments are discussed. 
It is argued that both short and long baseline accelerator neutrino 
experiments are needed in 
order to decide between this solution and other oscillation scenarios.
\end{abstract} 
\vspace*{1cm}
{\sl to be published in Phys. Lett. B}  


\bigskip 
keywords: pseudo-Dirac, neutrino, oscillation, oscillations, sterile, mass  
\end{center} \vfill 
\titlefoot{CERN, EP Division, CH-1211 Gen\`eve 23,\quad Tel. +41-22-7678564, 
\quad Fax +41-22-7679070,\\ \quad e-mail Achim.Geiser@cern.ch \label{CERN}}
\vfill
\end{titlepage}

\hoffset -0.7in 
\textwidth 6.0in 
\textheight 9.0in 
\normalsize 
\pagenumbering{arabic}
%
%

\section{Introduction}

Currently there is a wealth of unexplained phenomena in neutrino physics which
can be interpreted as indications for the existence
of neutrino oscillations. The solar neutrino problem \cite{SOLAR}, lacking a 
satisfactory astrophysical solution, is commonly attributed to the 
disappearance of electron (anti)neutrinos into some other neutrino type.
Similarly, the atmospheric neutrino anomaly \cite{ATMOS} is often interpreted 
as an indication of neutrino oscillations involving $\nu_\mu$ disappearance,
with or without partial reappearance as $\nu_e$.
The LSND experiment \cite{LSND} claims direct evidence for $\nu_\mu - \nu_e$ 
oscillations in a region which is partially unconstrained by other 
experiments. 
Furthermore, neutrinos are a prime candidate for a partial solution
to the missing dark matter problem \cite{DM} if at least one mass eigenstate
lies in the eV range.
These current indications are summarized in Fig. \ref{fig:solution}.

Within the framework of 3 generation neutrino mixing, all recent 
solutions \cite{solutions1}\cite{solutions2}\cite{solutions3} to this 
neutrino puzzle either discard some of the 
experimental evidence or have to accept a bad fit to part of the data.
Other solutions \cite{Caldwell} involve the existence of one additional light 
sterile neutrino, yielding at least two extra degrees of freedom 
(mass and mixing angle(s)).
Solutions involving more than one such neutrino \cite{Mirror}\cite{pseudoold} 
have often been discarded due to
bounds on the number of neutrino types suggested by Big Bang nucleosynthesis
\cite{Schramm}\cite{actstemix}.

In this paper a solution is presented based on the assumption that neutrinos
are of the pseudo-Dirac type, i.e. $m_{\rm Majorana} \ll m_{\rm Dirac}$
\cite{numass}. 
This involves the splitting of each neutrino generation into two almost
degenerate mass eigenstates, leading to neutrino-antineutrino oscillations
with maximal mixing, similar to oscillations in the $K^0-\bar K^0$ system.
Since oscillations do not change the handedness, left-handed 
neutrinos (right handed antineutrinos) will transform into left-handed 
antineutrinos (right handed neutrinos) which do not 
participate in weak interactions, and therefore appear sterile.

In contrast to earlier schemes of this kind \cite{pseudoold}, the neutrino
magnetic moments are assumed to be 0, therefore avoiding 
$\nu_L \leftrightarrow \nu_R$ transitions. Special emphasis is placed on the 
case of exactly maximal mixing, which minimizes matter effects.
Furthermore, the scheme is extended to include 3 generation flavour mixing, 
and applied to the most recent experimental data. 

With only one parameter (a mass splitting parameter chosen to be universal 
for simplicity) in addition to the 
standard 3 generation mixing, {\sl all} experimental indications for
neutrino oscillations can be fully explained in the context of a 
hierarchical mass and mixing pattern, analogous to the 
quark sector. In addition, this solution
suggests a substantial neutrino contribution to the missing
dark matter. A mechanism through which the Big Bang nucleosynthesis bounds
mentioned earlier can be evaded is discussed in the appendix.

\section{The pseudo-Dirac neutrino formalism}

The most general neutrino mass term (Dirac-Majorana mass term) that can be 
added to the standard model Lagrangian leads, for one neutrino generation, 
to the (Majorana) mass eigenstates \cite{numass}\cite{Majorana}
$$ m_{I,II} = | \frac{1}{2} \left[ (m_L+m_R) \pm \sqrt{(m_L-m_R)^2 + 4 m_D^2} \right] | $$
where $m_D$ is the Dirac mass, and $m_L, m_R$ are the left and right handed
Majorana masses, and to a mixing angle 
$$ \tan 2\theta = 2m_D / (m_R-m_L) $$
between these two states.
 
The standard Dirac case is obtained by setting $m_L=m_R=0$, 
while $m_D=0$ leads to pure Majorana neutrinos. 
The limit $m_R \gg m_D,\ m_L = 0\ (\theta = m_D/m_R \ll 1)$
yields the mass eigenstates 
$m_I = m_D^2/m_R,\ m_{II} \simeq m_R$ which corresponds to the well-known
see-saw model \cite{seesaw}, where $m_D$ is assumed to be of the order of the quark
and charged lepton masses, and $m_R$ is a large mass of the order of some 
unification scale. This forces the mass of the ``left-handed'' neutrino 
($m_I$) to be small, and makes the ``right-handed'' neutrino ($m_{II}$) very
heavy, therefore decoupling it from interactions in the accessible
energy range.

Finally, the limit $m_D \gg m_L,m_R\ (\theta \simeq 45^o)$ yields almost 
degenerate mass eigenstates $m_{I,II} \simeq m_D$, with close to maximal
mixing arising in a ``natural'' way. This case, which is known 
as the pseudo-Dirac case, will be pursued further here. 
Taking $m_L=m_R=\delta m$ for simplicity one obtains
$$ m_{I,II} = m_D \pm \delta m,  \qquad \theta = 45^o .$$
In this case, the mass eigenstates can be written in terms of the 
neutrino/antineutrino eigenstates of the weak interactions as
$$ |\nu_I> = \frac{1}{\sqrt{2}} (|\nu > + |\bar \nu >), \qquad 
   |\nu_{II}> = \frac{1}{\sqrt{2}} (|\nu > - |\bar \nu >), $$
a pattern which is similar to the one in the $K^0 - \bar K^0$ meson system. 
As in the $K^0$ case, this yields maximal oscillations between particle
and antiparticle states if pure neutrinos or antineutrinos are produced.
Since, up to a small mass effect, neutrinos are emitted left-handed 
and oscillations cannot change the handedness (angular momentum conservation)
they will oscillate into left-handed antineutrinos which appear sterile.
The same is true for right-handed antineutrinos oscillating into right-handed
neutrinos. Thus
phenomenologically one obtains active-sterile neutrino oscillations with
$$\Delta m^2 = (m_D + \delta m)^2 - (m_D - \delta m)^2 = 4m_D\delta m$$  
and maximal mixing (sin$^2$2$\theta$ = 1).

This can easily be generalized to 3 generations. Neglecting CP violation, 
the resulting mixing matrix is a real $6 \times6$ matrix in principle. 
Here, we simplify it with the hypothesis
that the oscillations between different generations and the active-sterile
oscillations decouple, thus yielding the standard 3 generation mixing matrix
$U_{\alpha i}$ ($\alpha=e,\mu,\tau,\ i = 1,2,3$), equivalent to the CKM matrix
\cite{CKM}, plus maximal active-sterile 
mixing for each generation

$$ |\nu_\alpha > = U_{\alpha i} \frac{1}{\sqrt{2}} (|\nu_{i,I} > + |\nu_{i,II} >), \qquad
  |\bar \nu_\alpha > = U_{\alpha i} \frac{1}{\sqrt{2}} (|\nu_{i,I} > - |\nu_{i,II} >).$$ 

The above discussion remains valid to good approximation for $m_L \neq m_R$
as long as both terms are very small, and in particular for $m_L=0$.
Since $m_R$ mass terms can be introduced ad hoc as bare mass 
terms without destroying gauge invariance, no attempt is made at this stage 
to clarify the origin of the Majorana masses.
Phenomenologically, two cases are considered:
(1) same $\delta m_i = \varepsilon \times M$ for each generation, where 
$M=1$ eV is an arbitrary fixed reference mass,
 and  
(2) mass dependent splitting $\delta m_i = \varepsilon \times m_{Di}$.
The mass pattern generated by these scenarios is therefore given by
$m_{Di} \simeq m_1, m_2, m_3$ for the 3 generations, and by 
the single additional
parameter $\varepsilon \ll 1$ fixing the mass fine splittings.
   

\section{The neutrino oscillation solution}

To find a simple general solution to the neutrino oscillation puzzle
described in the introduction the following further assumptions are made,
in analogy to the quark sector:
\begin{itemize}
\item The mass structure is hierarchical, i.e. $m_1 \ll m_2 \ll m_3$.
\item Mixing between adjacent generations is small ($\theta_{12}$ and $\theta_{23}$) 
and mixing between far generations is negligible ($\theta_{13} \simeq 0$).
Flavour eigenstates can therefore be approximately identified with mass 
eigenstates ($1\simeq e$, $2\simeq \mu$, $3\simeq\tau$). 
\end{itemize}

Including $\varepsilon$ this yields a total of six free parameters 
($m_i (i=1,2,3), \varepsilon, \theta_{12}, \theta_{23}$). 

The $\Delta m^2$ values 
relevant for oscillations are $\Delta m^2_{12}, \Delta m^2_{23}$ for
conventional mixing, and $\Delta m^2_{I/II,i} = 4 m_i M \varepsilon$ for case (1)
($\Delta m^2_{I/II,i} = 4 m_i^2 \varepsilon$ for case (2)) for active-sterile oscillations.  
In the simplified case of only two contributing flavour and mass eigenstates, the 
probability of a neutrino of type $\alpha$ to oscillate into type $\beta$
is given by the well known formula
$$P_{\alpha \beta} = \sin^2 2\theta_{\alpha\beta}\ \sin^2(1.27\ \Delta m^2_{\alpha\beta}\ L/E)$$
where L is the distance travelled in km, E is the neutrino energy in GeV,
and $\Delta m^2_{\alpha\beta}$ is expressed in eV$^2$.
 
\begin{figure}
\begin{center}
\epsfig{file=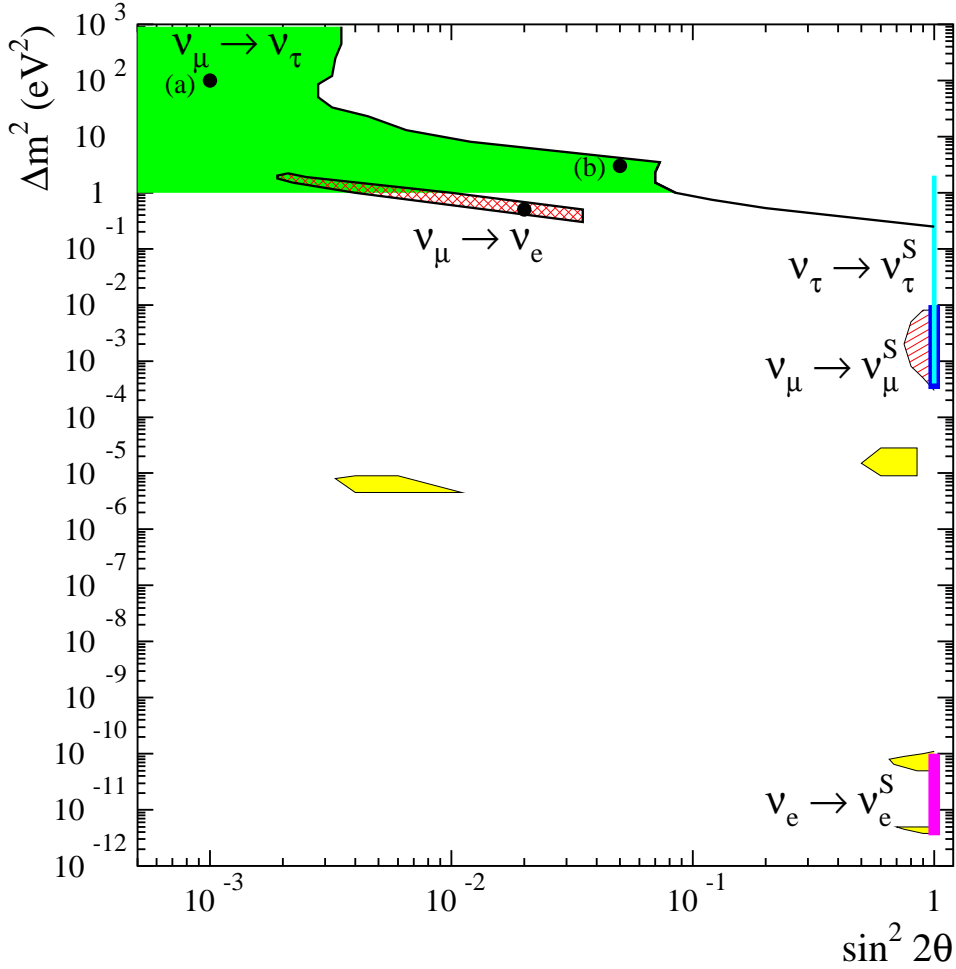,width=140mm}
\caption {Current indications for neutrino oscillations in the 
          sin$^2$2$\theta$ vs. $\Delta m^2$ plane, assuming dominance
          of two flavour oscillations in each case. The figure shows
          the 90\% c.l.
          $\nu_\mu~\to~\nu_e$ region favoured by LSND (double hatched), 
          including constraints from BNL 776 and 
          Bugey, the Superkamiokande $\nu_\mu$ disappearance solution to
          the atmospheric neutrino problem (hatched), the 3 mutually
          exclusive 95\% c.l. solutions to the solar neutrino problem 
          (light shaded),
          and the 90\% c.l. limit on $\nu_\mu \to \nu_\tau$ (black line).
          Also shown is the parameter space of the proposed solutions for 
          $\nu_X \to \nu_X^S$ (bars at sin$^2 2 \theta = 1$), 
          $\nu_\mu \to \nu_\tau$ (dark shaded) and $\nu_\mu \to \nu_e$ 
          (double hatched, identical to LSND region).
          The reference solutions described in the text are also indicated
          (black circles).}
\label{fig:solution}
\end{center}
\end{figure}

In the following, a complete solution for the six free parameters is derived 
based on all known experimental evidence (table \ref{tab:solution}). 
A graphical representation of this solution is shown in Fig. 
\ref{fig:solution}. 

\subsection{LSND}

The LSND result \cite{LSND} is interpreted as direct $\nu_\mu - \nu_e$ 
oscillations.
The large $\Delta m^2$ region ($\Delta m^2 > 10$ eV$^2$) is excluded by 
NOMAD \cite{Jerusalem}, CCFR \cite{CCFR}, BNL 734 \cite{BNL734}, 
and BNL E766 \cite{BNL776}. The latter 
constrains also the island around $\Delta m^2 \sim 6$ eV$^2$ advocated by 
Caldwell \cite{Caldwell1}.
The remaining region ($\Delta m^2 \le 2$ eV$^2$) is fully consistent with, 
and even slightly favoured by, the recent KARMEN1 limits \cite{KARMEN} 
partially based on a small ($\sim 1 \sigma$) positive result.
At large sin$^2 2 \theta$ the result is constrained by reactor experiments,
including G\"osgen \cite{Goesgen}, Krasnoyarsk \cite{Krasnoyarsk}, 
Chooz \cite{Chooz} and Bugey \cite{Bugey}.
Note that the Bugey result, which yields the most stringent limit, is also
based on a small (up to 1.5 $\sigma$) positive effect \cite{Bugey}, 
fully consistent with the limits
from the other reactor experiments. Inspired by this effect
we choose a reference solution $\Delta m^2_{12} \sim 0.5$ eV$^2$, sin$^2 2 
\theta_{12} \sim 2 \times 10^{-2}$. However, 
no significant lower constraint on sin$^2 2 \theta$ can be derived from the
Bugey result, allowing sin$^2 2 \theta_{12}$ to be as small as 
$2 \times 10^{-3}$ 
for $\Delta m^2_{12} \sim 2$~eV$^2$.


\subsection{Solar neutrinos}

The solar neutrino problem is interpreted as $\nu_e -\nu_e^S$ oscillations
with sin$^2$2$\theta_{eS}$~=~1.
This yields an average expected suppression factor of 0.5, slightly reduced 
further by a O(1\%) correction from $\nu_e -\nu_\mu$ oscillations.
Energy dependent oscillations have to be introduced in order to describe the 
observed suppression factors \cite{MSWsolar} with respect to the standard 
solar model \cite{Bahcall}
of 0.51 $\pm$ 0.05 for the Gallium experiments \cite{GALLEX},  
0.27 $\pm$ 0.02 for the Homestake experiment \cite{Homestake},
and 0.39 $\pm$ 0.03 for the water \v{C}erenkov detectors 
\cite{Kamiokande}\cite{SuperK}.
Since the MSW solution \cite{MSWsolar}\cite{MSW} is irrelevant for maximal 
mixing, 
the vacuum solution \cite{MSWsolar}\cite{Vacuum} is the only possible one. 
This yields $\Delta m^2_{I/II,1} \sim 5 \times 10^{-12} - 10^{-10}$ eV$^2$.

\subsection{Atmospheric neutrinos}

The atmospheric neutrino anomaly is interpreted as $\nu_\mu -\nu_\mu^S$
oscillations with sin$^2$2$\theta_{\mu S}$~=~1. A small effect from
$\nu_\mu - \nu_e$ and $\nu_\mu -\nu_\tau$ oscillations should be added.
The $\Delta m^2$ for $\nu_e$ disappearance 
($\Delta m^2_{I/II,1}$)  
is too small to have any influence,
so electron-like events are expected to be slightly enhanced by 
$\nu_\mu - \nu_e$ oscillations without L/E dependence. On the other hand, 
muon-like events are depleted mainly by $\nu_\mu -\nu_\mu^S$ oscillations, 
with a small L/E independent suppression from $\nu_\mu - \nu_e$, and another
small but potentially complicated contribution from $\nu_\mu - \nu_\tau$ 
followed by $\nu_\tau - \nu_\tau^S$. Using the LSND reference solution quoted 
above, an L/E independent suppression would imply a value of 
$R=\frac{\mu/e_{measured}}{\mu/e_{expected}}$ of 0.49 or lower 
(if $\nu_\mu - \nu_\tau$ is significant). The observed average suppression
of only about 0.63 \cite{SuperK} is therefore an indication for non-maximal 
$\nu_\mu$ suppression due to dependence on L/E. Indeed such a dependence is 
observed indirectly through a dependence on the zenith angle by Kamiokande
\cite{Kamatmo} and Superkamiokande \cite{Supatmo}. 
Reinterpretation of the Superkamiokande result in terms of  maximal 
$\nu_\mu -\nu_\mu^S$ oscillations suggests a value of 
$\Delta m^2_{I/II,2}$ in the range $3 \times 10^{-4} - 10^{-2}$ eV$^2$
\cite{Volkas1}. The detailed consideration of matter effects can potentially
induce a slight upward shift of this range \cite{Volkas2}.  
Such low $\Delta m^2$ values can also explain why the suppression is stronger 
for the water \v{C}erenkov detectors \cite{Kamatmo}\cite{Supatmo}
than for some of the iron detectors \cite{Frejus} which have a 
higher energy threshold.
Experiments looking for upward going muons yield inconclusive results so far 
\cite{IMBup}\cite{MACRO}, but are not inconsistent 
with the $\nu_\mu -\nu_\mu^S$ interpretation.

\subsection{Dark matter}

The potential contribution of neutrinos to hot dark matter is given by
$\Omega_\nu = \Sigma m_\nu / (92 h^2\ {\rm eV})$ \cite{Srednicki}, 
where $\Omega_\nu$ is the neutrino energy density parameter expressed as
the fraction of the critical density $\Omega = 1$ needed to close the 
universe, $h$ is the dimensionless renormalized Hubble constant 
($0.5 < h < 0.85$ \cite{Lang}) and $\Sigma m_\nu$ is the sum over the masses 
of all neutrino types. 
For pseudo-Dirac neutrinos each neutrino flavour enters this sum twice if
its sterile counterpart reaches thermal equilibrium in the early universe, 
and at least once otherwise.
In the hierarchical mass pattern advocated here this 
contribution arises mainly from $\tau$ neutrinos \cite{HARARI}. 
No direct evidence for $\nu_\mu - \nu_\tau$ oscillations exists so far 
in the context of the pseudo-Dirac solution.
Constraints are imposed by the exclusion regions obtained
by accelerator experiments 
\cite{Jerusalem}\cite{CHORUS}\cite{E531}\cite{LIMITS}. 
Two typical solutions close to the range
accessible by current experiments and relevant to the dark matter problem 
could be (a)~$\Delta m^2_{23}~=~100$~eV$^2$ ($m_3~=~10$~eV) 
with sin$^2$2$\theta_{23}~=~10^{-3}$,
and (b) $\Delta m^2_{23} = 3$ eV$^2$ ($m_3 \simeq 2$ eV) with 
sin$^2$2$\theta_{23}~=~5\times10^{-2}$.
Solution (a), with its very small mixing, would have virtually no effect on 
any of the current measurements with positive indications for oscillations, 
but yields a large hot dark matter component ($\Omega_\nu \sim 0.4$ for
$h \sim 0.7$) which is disfavoured by most
current dark matter scenarios \cite{DM}. The associated $\Delta m^2_{23}$ 
could however be lowered to any value larger than 1 eV$^2$ 
(mass hierarchy limit), corresponding to $\Omega_\nu > 0.03$,
without significantly changing the conclusions. 
Alternatively, it could be raised up to the extreme limit from pure 
hot dark matter models of about (30 eV)$^2$ \cite{HARARI}.  
Solution (b) would yield a small but 
significant contribution of about 5\% to the atmospheric neutrino suppression,
with only a moderate hot dark matter contribution  
($\Omega_\nu \sim 0.1$ for $h \sim 0.7$).

\subsection{The mass fine splitting}
\label{sect:split}

The $\Delta m^2$ for $\nu_\tau - \nu_\tau^S$ oscillations is essentially 
unconstrained by current data. Since $m_2\sim~1$~eV (LSND), the 
atmospheric neutrino solution 
yields $\varepsilon \sim 10^{-4} - 2 \times 10^{-3}$ for both case (1)
and  case (2).
Folding this with the $\Delta m^2_{23}$ parameter space spanned by 
$\nu_\mu - \nu_\tau$ solutions (a) and (b), 
$\Delta m^2_{I/II,3}$ could be anywhere in the range $\sim 10^{-3} -1$
eV$^2$.
In combination with $\nu_\mu - \nu_\tau$ oscillations this could yield a
small contribution to the angular effect observed in the atmospheric
neutrino anomaly.

Finally, the above $\varepsilon$ value applied to the solar neutrino result
predicts $m_1$ in the range $10^{-9} - 10^{-6}$ eV for case (1), 
and $10^{-5} - 10^{-3}$ eV for case (2). This mass and mixing pattern
is in agreement with the nonobservation of neutrinoless 
double $\beta$ decay \cite{doubleB}.

\begin{table}[h] 
  \caption{Parameters of the pseudo-Dirac neutrino solution for
           $\varepsilon=10^{-4}-2\times 10^{-3}$ 
           (cases (1) and (2) of section \ref{sect:split} combined).
           $m$ refers to the Dirac mass of each neutrino generation and 
           $\Delta m^2$ to either the mass fine splitting
           inside each generation, or the mass difference squared between 
           generations where appropriate. sin$^2 2 \theta$ refers to the 
           associated mixing angle.
           Only the entries in the $m$ and sin$^2 2 \theta$ columns are
           independent parameters, $\Delta m^2$ being fixed by $m$ and
           $\varepsilon$.}  
  \label{tab:solution}
  \begin{center}
    {\normalsize
      \begin{tabular}{|l|c|c|c|}
        \hline
        generation & $m$ (eV) & $\Delta m^2$ (eV$^2$) & sin$^2 2 \theta$ \cr
        \hline
        1 (``$e$'') & $10^{-9}-10^{-3}$ & $5\times 10^{-12} - 10^{-10}$ & 1 \cr
        \hline
        2 (``$\mu$'')& $0.5-1.7$ & $3\times 10^{-4} - 10^{-2}$ & 1 \cr
        \hline
        3 (``$\tau$'')& $1-30$ & $4\times 10^{-4} - 2$ & 1 \cr
        \hline
        $1 \leftrightarrow 2$ & $-$ & $0.3 - 3$ & $0.002-0.035$ \cr
        \hline
        $2 \leftrightarrow 3$ & $-$ & $1 - 30$ & $< 0.07$ \cr
              &     & or $30-900$ & $< 0.0035$ \cr
        \hline
        $1 \leftrightarrow 3$ & $-$ & $1-900$ & negligible \cr
        \hline
      \end{tabular}
      }
  \end{center}
\end{table}
\vspace*{-0.5cm}

\section{Alternative solutions}

Only two of the critical experimental inputs are so far not cross-checked 
independently by at least one other experiment: The LSND result \cite{LSND}, 
and the Homestake result \cite{Homestake} for solar neutrinos. 
The former fixes the $\nu_\mu - \nu_e$
oscillation parameters and, if disproven, would essentially remove any
lower constraint on $\Delta m^2_{12}$. Comparison of the latter to the
other solar neutrino experiments yields the main evidence for the 
energy-dependence of the solar neutrino deficit. A 30\% increase of the 
Homestake neutrino flux measurement (not compatible with the quoted error), 
and a reduction of the 
predicted $B^8$ solar neutrino flux by about 20\% (which is inside the
currently discussed error margin \cite{Boron}) would make all solar neutrino
measurements compatible with a flat suppression factor of about 0.5
(see also \cite{solutions1}\cite{solutions3}). This would
remove the constraint on $\Delta m^2_{I/II,1}$ which,
ignoring Big Bang nucleosynthesis implications, 
could then be as large as $10^{-4}$~eV$^2$ 
(bounded by its potential effect on the atmospheric neutrino anomaly). 
In order to fix the parameters of the pseudo-Dirac model
the verification of LSND by KARMEN2 \cite{KARMEN} and later
by other experiments \cite{numunue} is therefore crucial, and a verification 
of the energy dependence of the solar neutrino suppression, e.g. by Borexino 
\cite{Borexino}, is highly desirable.
Note that the azimuthal dependence of the atmospheric neutrino suppression,
which is observed by only one type of experiment, is {\sl not} an 
a priori independent experimental input in this context, but rather a 
prediction of the model based on the observed non-maximal
$\nu_\mu$ suppression.

\section{Predictions for current and future experiments}

The solution of the neutrino oscillation puzzle proposed in this paper can be 
checked by several current and future experiments.

Since the solar neutrino problem is solved via $\nu_e$ disappearance into
sterile neutrinos, SNO \cite{SNO} should not observe an enhancement of the 
NC/CC ratio as it would in the case of $\nu_e \to \nu_\mu$.
No day/night effect should be present, but a small seasonal effect might
be expected (vacuum solution).

The atmospheric neutrino anomaly should be confirmed, including the 
azimuthal dependence of the $\mu/e$ suppression. Since it is mainly due to 
$\nu_\mu$ disappearance into sterile neutrinos, the muon deficit (rather
than electron excess) should be confirmed, and the $\nu_\mu$ NC/CC ratio 
should not be 
enhanced. Experiments looking at upward
going muons \cite{MACRO} should also see an effect, and current reactor 
experiments
testing $\nu_\mu -\nu_e$ oscillations \cite{Chooz}\cite{Palo} should continue 
to observe negative or marginal results. 
Long baseline accelerator experiments \cite{ICARUS}\cite{MINOS}\cite{OPERA}
should not observe a strong $\nu_\tau$ appearance effect, but
should observe maximal $\nu_\mu$ disappearance if the 
$\nu_\mu - \nu_\mu^S$ oscillations 
lie in the experimentally accessible range.
Unfortunately, the low $\Delta m^2$ range ($\Delta m^2 < 10^{-3}$~eV$^2$) 
allowed by Superkamiokande \cite{SuperK} can not be fully tested by any of 
the currently planned long baseline experiments. 

However, ICARUS \cite{ICARUS} and MINOS \cite{MINOS}, if sensitive enough, 
should observe energy independent $\nu_\mu - \nu_e$ 
oscillations, whose (small) rate should finally settle the value of 
sin$^2 2 \theta$ for the LSND result.
  
$\nu_\mu - \nu_\tau$ oscillations should be observed in short baseline
accelerator experiments within the next decade.
If not by the current experiments NOMAD and CHORUS \cite{CHORUS}, 
then by future more
sensitive experiments \cite{TOSCA} with a slightly lower $\Delta m^2$ 
threshold. $\nu_\mu - \nu_\tau$ solution (b) would also be within the 
reach of the long baseline $\tau$ appearance experiments \cite{ICARUS}\cite{OPERA}, and could possibly be confused with a signal at low $\Delta m^2$
in the atmospheric neutrino
region. Therefore both short {\sl and} long baseline experiments are needed
to unambigously confirm or rule out this oscillation scenario.
Since $m_3$ is of order eV or more (hierarchy), $\tau$ neutrinos 
should make up a significant
part of the missing dark matter.

If $\nu_\mu -\nu_\tau$ oscillations would be discovered with sizeable
mixing, $\nu_\tau - \nu_\tau^S$
disappearance could be verified in the long term by comparing the $\tau$ production 
rates in short, medium, and long baseline experiments. Some corrections to
the atmospheric neutrino spectrum might also be observable.

Lastly, if this pseudo-Dirac neutrino solution would be confirmed and some of
its parameters measured more precisely, the complicated interplay between 
lepton number violating active-sterile neutrino oscillations, matter effects, 
and Big Bang nucleosynthesis (see appendix) could be used to further constrain
the remaining parameters.  

\setcounter{secnumdepth}{0} 

\section{Appendix: Compatibility with Big Bang nucleosynthesis}

As is well known \cite{DHY}, 
the ratio of deuterium to hydrogen (D/H) and the fraction of He$^4$ in the 
universe (Y) is determined in the Big Bang model by the ratio of neutrons to 
protons at the time of the ``weak freeze out'', i.e. the time when the 
reactions $n + e^+ \leftrightarrow p + \bar\nu_e$ and 
$p + e^- \leftrightarrow n + \nu_e$ proceed too slowly 
compared to the expansion rate of the universe to keep n/p at its 
thermal equilibrium value.
The effective number of light neutrino flavours $N_\nu$ enters as a 
contribution to the 
energy density, which in turn influences the expansion rate.
From this effect, a typical upper limit 
derived from current Big Bang nucleosynthesis (BBN) models is 
$N_\nu < 3.1$ @ 95\% c.l. \cite{DHY} with a central value of about 2.3,
only marginally compatible with the standard
model $N_\nu = 3$.

\par\goodbreak

Active-sterile neutrino mixing can influence this rate in two ways:
If active-sterile oscillations occur during thermal equilibrium, i.e.
significantly before weak freeze out, 
the sterile neutrinos will make an additional
contribution to the energy density at freeze out, and therefore 
increase the expansion rate, effectively behaving like
additional light neutrino generations. On the other hand,
depletion of $\nu_e$ due to $\nu_e \to \nu_e^S$ during freeze-out  
reduces the reaction rate for n/p interchange. In both cases, 
the predicted D/H and Y values will be larger than in the standard
$N_\nu = 3$ case, and therefore even less compatible with the 
observed values which yield the limit quoted above.

The $\Delta m^2$ of $10^{-10}$ eV$^2$ or lower for 
vacuum $\nu_e - \nu_e^S$ oscillations yields an oscillation time scale which 
is significantly longer than the freeze out time scale. The oscillation
effect is thus negligible for practical purposes \cite{Schramm}. 
Both $\nu_\mu - \nu_\mu^S$ and
$\nu_\tau - \nu_\tau^S$ oscillations approximately fall in the mass range 
$10^{-3} - 10^{-1}$ eV$^2$ for which the BBN implications are discussed in 
ref. \cite{actstemix}. Ignoring non standard model effects such as lepton
number violation, each 
of these contribute 1 unit to $N_\nu$ if 
$\Delta m^2$ is at the upper edge of this range, and slightly less than one
unit if $\Delta m^2$ is of order $10^{-3} - 10^{-4}$ \cite{Schramm},
yielding $N_\nu \sim 4.6-5$. 

However, matter effects induced by a significant initial lepton number 
asymmetry ($10^{-3}$ or larger) can almost {\sl fully suppress} the 
$\nu_\mu - \nu_\mu^S$ and $\nu_\tau - \nu_\tau^S$ oscillations during thermal
equilibrium \cite{Foot1}, therefore avoiding any contribution to the 
effective number of neutrinos.

Moreover, such an asymmetry can be created by the (lepton number 
violating) active-sterile oscillations {\sl themselves} \cite{Foot2}, 
if small inter-generation oscillations (i.e. $\nu_\tau - \nu_\mu^S$) are 
added.   
The quantitative treatment of the $\nu_\mu - \nu_\mu^S$ case in ref. 
\cite{Foot3} is qualitatively also applicable to 
$\nu_\tau - \nu_\tau^S$ oscillations.
Interestingly, the parameters needed to create a lepton number asymmetry which
suppresses $\nu_\mu - \nu_\mu^S$ and 
$\nu_\tau - \nu_\tau^S$ oscillations until they become irrelevant for BBN
largely overlap with the parameter space suggested by the pseudo-Dirac 
solution (table \ref{tab:solution}).
For example, taking $\Delta m_{I/II,2}^2 = 10^{-3}$ eV$^2$, 
$\Delta m_{12}^2 = 1$ eV$^2$ ($\varepsilon = 3 \times 10^{-3}$ for case (1)) 
 and adding  
$\nu_\tau - \nu_\mu^S$ or $\nu_\tau - \nu_e^S$ oscillations with
sin$^2 2 \theta \sim 10^{-6}$, which have a negligible effect on the 
phenomenology of the pseudo-Dirac model, almost full $\nu_X - \nu_X^S$ 
suppression ($X=\mu,\tau$)
is obtained for $\Delta m_{23}^2 > 3$ eV$^2$, which is just
what is required from the hierarchy assumption.

Finally, with a similar set of parameters, the ``reprocessing'' of the initial
lepton number asymmetry into an effective excess of $\nu_e$ over $\bar{\nu_e}$
and a corresponding asymmetry in the $n \to p$ and $p \to n$ reaction rates 
during BBN can induce a {\sl reduction} of the effective number of neutrinos
by as much as 0.5 \cite{Foot4}. This yields the lower bound $N_\nu \geq 2.5$,
compatible with the current limit.
Furthermore, the systematic error on the experimental input for Y and D/H is
a matter of ongoing discussion \cite{nnulimit}, which could contribute to
loosening the BBN constraint.

In conclusion, the apparent conflict between the BBN limit and 3 additional 
sterile neutrinos can be evaded through the lepton number violating 
(i.e. non standard model) effects caused by active-sterile neutrino 
oscillations.



\section{Acknowledgements}

Constructive discussions with L. Camilleri, Ch. Cardall, J. Ellis, R. Foot, 
E.~Nagy, U.~Stiegler and R.~Volkas are gratefully acknowledged.  


%

\begin{thebibliography}{99}
\bibitem{SOLAR} see e.g. V. Castellani et al., Phys. Rep. {\bf 281} (1997) 309.
\bibitem{ATMOS} see e.g. T.K. Gaisser, in Neutrino '96, Proceedings of the 
17th International Conference on Neutrino Physics and Astrophysics, Helsinki, 
Finland, June 13, 1996,
ed. by K. Enqvist, K. Huitu and J. Maalampi, World Scientific (1997), p. 211.
\bibitem{LSND} LSND Collaboration, C. Athanassopoulos et al., 
               Phys. Rev. {\bf C54}, 2685 (1996); \\
               LSND Collaboration, C. Athanassopoulos et al., 
               Phys. Rev. Lett. {\bf77}, 3082 (1996); \\
               LSND Collaboration, C. Athanassopoulos et al., LA-UR-97-1998, 
               UCRHEP-E191, submitted to Phys. Rev. {\bf C}; 
               UCRHEP-E197, submitted to Phys. Rev. Lett. 
\bibitem{DM} see e.g. J. Primack and A. Klypin, Nucl. Phys. Proc. Suppl. 
{\bf 51 B} (1996) 30.  
\bibitem{solutions1} A. Acker and S. Pakvasa, 
Phys. Lett. {\bf B 397} (1997) 209; \\
P.F. Harrison, D.H. Perkins, and W.G. Scott, 
Phys. Lett. {\bf B 349} (1995) 137; Phys. Lett. {\bf B 349} (1997) 186; \\
E. Torrente-Lujan, Phys. Lett. {\bf B 389} (1996) 557. 
\bibitem{solutions2}C.Y. Cardall and G.M. Fuller, 
Phys. Rev. {\bf D 53} (1996) 4421; \\
K.S. Babu, J.C. Pati, and F. Wilczek, Phys. Lett. {\bf B 359} (1995) 351; \\
G.L. Fogli, E. Lisi, and D. Montanino, Phys. Rev. {\bf D 49} (1994) 3626;
Phys. Rev. {\bf D 54} (1996) 2048;
G.L. Fogli, E. Lisi, and G. Scioscia, Phys. Rev. {\bf D 52} (1995) 5334;
Phys. Rev. {\bf D 56} (1997) 3081;
G.L. Fogli et al., Phys. Rev. {\bf D 56} (1997) 4365; \\
H. Minakata, Phys. Lett. {\bf B 356} (1995) 61; \\
S.M. Bilenky et al., Phys. Lett. {\bf B 356} (1995) 273; 
Phys. Rev. {\bf D 54} (1996) 1881.
\bibitem{solutions3}
G. Conforto et al., Astropart. Phys. {\bf 5} (1996) 147; \\
S. Barshay and P. Heiliger, Astropart. Phys. {\bf 6} (1997) 323.
\bibitem{Caldwell} D.O. Caldwell and R.N. Mohapatra, Phys. Rev. {\bf D 48}
(1993) 3259; Phys. Rev. {\bf D 50} (1994) 3477; \\
J.T. Peltoniemi and J.W.F. Valle, Nucl. Phys. {\bf B 406} (1993) 409; \\
J.T. Peltoniemi, D. Tommasini and J.W.F. Valle, Phys. Lett. {\bf B 298} (1993) 383; \\ 
H. Minakata, Phys. Rev. {\bf D 52} (1995) 6630; \\
D. Suematsu, Phys. Lett. {\bf B 392} (1997) 413; \\
Z.G. Berezhiani and R.B. Mohapatra, Phys. Rev. {\bf D 52} (1995) 6607; \\
J.J. Gomez-Cadenas and G.M. Gonzalez-Garcia, Z. Phys. {\bf C 71} (1996) 1996.
\bibitem{Mirror} R. Foot and R.R. Volkas, Phys. Rev. {\bf D 52} (1995) 6595;\\
                 R. Foot, Mod. Phys. Lett. {\bf A 9} (1994) 169;\\
                 J.P. Bowes and R.R. Volkas, hep-ph/9804310, to appear in
                 J. Phys. {\bf G}.   
\bibitem{pseudoold} M. Kobayashi, C.S. Lim and M.M. Nojiri, Phys. Rev. Lett.
{\bf 67} (1991) 1685; \\
C. Giunti, C.W. Kim and U.W. Lee, Phys. Rev. {\bf D 46} (1992) 3034.
\bibitem{Schramm} X. Shi, D.N. Schramm and B.D. Fields, Phys. Rev. {\bf D 48} (1993) 2563; \\
K. Enqvist, K. Kainulainen and M. Thomson, Nucl. Phys. {\bf B 373} (1992) 498.
\bibitem{actstemix} C.Y. Cardall and G.M. Fuller, Phys. Rev. {\bf D 54} (1996) 
1260.
\bibitem{numass} L. Wolfenstein, Nucl. Phys. {\bf B 186} (1981) 147; \\
S.M. Bilenky and B.M. Pontecorvo, Yad. Fiz. {\bf 38} (1983) 415 
(Sov. J. Nucl. Phys. {\bf 38} (1983) 248).
\bibitem{Majorana} G. Gelmini and E. Roulet, Rep. Prog. Phys. {\bf 58}
(1995) 1207.
\bibitem{seesaw} J. Ellis and M. Karliner, Phys. Lett. {\bf B 213} (1988) 73;\\
               D.B. Kaplan and A. Manohar, Nucl. Phys. {\bf B 310} (1988) 527. 
\bibitem{CKM} N. Cabibbo, Phys. Rev. Lett {\bf 10} (1963) 531; 
M. Kobayashi and T. Maskawa, Progr. Th. Phys. {\bf 49} (1973) 652; \\
F.J. Gilman, K. Kleinknecht and B. Renk, in Review of Particle Physics,
Phys. Rev. {\bf D 54} (1996), p. 94.       
\bibitem{Jerusalem} V. Valuev (NOMAD Collaboration), proceedings of the 
International Europhysics Conference on High Energy Physics, Jerusalem,
Israel, 19-26 August 1997, to be published. 
\bibitem{CCFR} CCFR Collaboration, A. Romosan et al., Phys. Rev. Lett. 
{\bf 78} (1997) 2912.
\bibitem{BNL734} L.A. Ahrens et al., Phys. Rev. {\bf D 31} (1985) 2732.
\bibitem{BNL776} L. Borodowsky et al.,Phys. Rev. Lett. {\bf 68} (1992) 274.
\bibitem{Caldwell1} D.O. Caldwell, in Neutrino '96 \cite{ATMOS}, p. 182.
\bibitem{KARMEN} J. Kleinfeller (KARMEN Collaboration), 
Nucl. Phys. Proc. Suppl. {\bf B 48} (1996) 207; \\
KARMEN Collaboration, B. Armbruster et al., Phys. Rev. {\bf C 57} (1998) 3414, 
and references therein; \\
K. Eitel (KARMEN Collaboration), CERN Particle Physics seminar, Febr. 17, 1998.
\bibitem{Goesgen} G. Zacek et al., Phys. Rev. {\bf D 34} (1986) 2621.
\bibitem{Krasnoyarsk} G.S. Vidyakin et al., JETP Lett. {\bf 59} (1994) 237.
\bibitem{Chooz} CHOOZ Collaboration, M. Apollonio et al., Phys. Lett. 
{\bf B 420} (1998) 397.
\bibitem{Bugey} B. Achkar et al., Nucl. Phys. {\bf B 434} (1995) 503.
\bibitem{MSWsolar} N. Hata and P. Langacker, Phys. Rev. {\bf D 56} (1997) 6107.
\bibitem{Bahcall} J.N. Bahcall and M.H. Pinsonneault, Rev. Mod. Phys. {\bf 67}
(1995) 781.
\bibitem{GALLEX} GALLEX Collaboration, P. Anselmann et al.,
Phys. Lett. {\bf B 327} (1994) 377;  
Phys. Lett. {\bf B 342} (1995) 440;
W. Hampel et al., Phys. Lett. {\bf B 388} (1996) 384; \\
SAGE Collaboration, J.N. Abdurashitov et al.,   
Phys. Lett. {\bf B 328} (1994) 234;  
Nucl. Phys. {\bf B 38} (1995) 60;  
Phys. Rev. Lett. {\bf 77} (1996) 4708.
\bibitem{Homestake} B.T. Cleveland et al., Nucl. Phys. B (Proc. Suppl.) 
{\bf 38} (1995) 47, and references therein.
\bibitem{Kamiokande} Kamiokande Collaboration, Y. Fukuda et al., 
Phys. Rev. Lett. {\bf 77} (1996) 1683. 
\bibitem{SuperK} 
Y. Totsuka (Superkamiokande results), Proceedings
of the Int. Lepton-Photon Symposium, Hamburg 1997 (in press). 
\bibitem{MSW} L. Wolfenstein, Phys. Rev. {\bf D 17} (1978) 2369; 
Phys. Rev. {\bf D 20} (1979) 2634; \\
S.P. Mikheyev and A.Y. Smirnov, Sov. J. Nucl. Phys. {\bf 42} (1985) 913;
Nuovo Cimento {\bf 9C} (1986) 17.
\bibitem{Vacuum}
      P.I. Krastev and S.T. Petcov, Phys. Rev. Lett. {\bf 72} (1994);
      Phys. Rev. {\bf D 53} (1996) 1665.
\bibitem{Kamatmo} Kamiokande Collaboration, Y. Fukuda et al., Phys. Lett. {\bf B 335} (1994) 237; \\
IMB Collaboration, R. Becker-Szendy et al., Phys. Rev. {\bf D 46} (1992) 237.
\bibitem{Supatmo} Super-Kamiokande Collaboration, Y. Fukuda et al., 
hep-ex/9803006 and hep-ex/9805006.
\bibitem{Volkas1} R. Foot, R.R. Volkas and O. Yasuda, Phys. Rev. {\bf D} (1998)
R1345.
\bibitem{Volkas2} R. Foot, R.R. Volkas and O. Yasuda, TMUP-HEL-9801,UM-P-98/04,RCHEP-98/01 (1998).
\bibitem{Frejus} NUSEX Collaboration, M. Aglietta et al., Europhys. Lett. {\bf 8} (1989) 611; \\
Fr\'ejus Collaboration, K. Daum et al., Z. Phys. {\bf C 66} (1995) 417; \\
Soudan II Collaboration, W.W.M. Allison et al., Phys. Lett. {\bf B 391} (1997) 491.  
\bibitem{IMBup} P. Lipari, M. Lusignoli and F. Sartogo, Phys. Rev. Lett.
{\bf 74} (1995) 4384.
\bibitem{MACRO} MACRO Collaboration, M. Ambrosio et al., INFN/AE-97/55, 
submitted to Phys. Rev. {\bf D}.
\bibitem{Srednicki} M. Srednicki, in Review of Particle Physics,
Phys. Rev. {\bf D 54} (1996), p. 116.       
\bibitem{Lang} K.R. Lang et al., in Review of Particle Physics,
Phys. Rev. {\bf D 54} (1996), p. 66.       
\bibitem{HARARI} H. Harari, Phys. Lett. {\bf B 216} (1989) 413; \\
                 J. Ellis, J.L. Lopez and D.V. Nanopoulos, Phys. Lett. {\bf B 292} (1992) 189.
\bibitem{CHORUS} CHORUS Collaboration, E. Eskut et al., 
Phys. Lett.~{\bf B 424} (1998) 202; \\
NOMAD Collaboration, J. Altegoer et al., CERN-EP/98-57, to appear in Phys. Lett.~{\bf B};\\ 
CHORUS Collaboration, E. Eskut et al., CERN-EP/98-149,
submitted to Phys. Lett.~{\bf B}.
\bibitem{E531} E531 Collaboration, N. Ushida et al., Phys. Rev. Lett. {\bf 57} 2897 (1986).
\bibitem{LIMITS} 
CHARM-II Collaboration, M. Gruwe et al., 
Phys. Lett. {\bf B 309} (1993) 463; \\
CCFR Collaboration, K.S. McFarland et al., 
Phys. Rev. Lett. {\bf 75} (1995) 3993; \\
CDHS Collaboration, F. Dydak et al., 
Phys. Lett. {\bf B 134} (1984) 281.
\bibitem{doubleB} P. Vogel, in Review of Particle Physics,
Phys. Rev. {\bf D 54} (1996), p. 289, and references therein.       
\bibitem{Boron} J. Bahcall, M.H. Pinsonnault, in Neutrino '96 \cite{ATMOS},
p. 56. 
\bibitem{numunue} Boone letter of intent, E. Church et al., LA-UR-97-2120 
                  (nucl-ex/9706011) (1997); \\ 
                  Letter of intent, ``Search for $\nu_\mu - \nu_e$ oscillations
                  at the CERN PS'', CERN-SPSC/97-21, SPSC/I 216, (1997).
\bibitem{Borexino} Borexino proposal, C. Arpesella at al., LNGS preprint 
(1991); \\
Borexino collaboration, G. Bellini et al, in TAUP '95 \cite{SNO}, p. 363. 
\bibitem{SNO} A. B. McDonald (SNO Collaboration), in TAUP '95, Proceedings of the 4th International Conference on Theoretical and
     Phenomenological Aspects of Underground Physics, Toledo, Spain, 1995, ed. by A. Morales, J. Morales, and J. A. Villar, [Nucl. Phys.
     B 48 (Proc. Suppl.)], p. 357; M. E. Moorhead, ibidem, p. 378.
\bibitem{Palo} Palo Verde proposal, F. Boehm et al., Caltech 1994;
               F. B\"ohm et al., Calt-63-721.
\bibitem{ICARUS} ICARUS Collaboration, P. Benetti et al., Nucl. Instr. Methods
                 {\bf A 327} (1993) 327; {\bf A 332} (1993) 332;
                P. Cennini et al., Nucl. Instr. Methods {\bf A 333} (1993) 567;
                 {\bf A 345} (1994) 230; {\bf A 355} (1995) 355; \\
                 ICARUS-CERN-Milano Collaboration, CERN/SPSLC 96-58, 
                 SPSLC/P 304, December 1996.
\bibitem{MINOS} MINOS Collaboration, Fermilab proposal P875; \\
                S.G. Wojcicki, in Neutrino '96 \cite{ATMOS}, p. 231.
\bibitem{OPERA} OPERA letter of intent, S. Shibuya et al., CERN-SPSC/97-24, SPSC/I 218, LNGS-LOI-8-97 (1997); \\
                Y. Suzuki (E362 (K2K) Collaboration), in Neutrino '96 
                \cite{ATMOS}, p. 237.
\bibitem{TOSCA} TOSCA letter of intent, CERN-SPSC/97-5, SPSC/I 213 (1997); \\
                R. A. Sidwell (COSMOS Collaboration) in Neutrino '96
                \cite{ATMOS}, p. 152.
\bibitem{DHY} K.A. Olive and D.N. Schramm, in Review of Particle Physics,
Phys. Rev. {\bf D 54} (1996), p. 109, and references therein.       
\bibitem{Foot1} R. Foot and R.R. Volkas, Phys. Rev. Lett. {\bf 75} (1995) 4350.
\bibitem{Foot2} R. Foot, M.J. Thomson and R.R. Volkas, Phys. Rev. {\bf D 53} 
(1996) 5349.
\bibitem{Foot3} R. Foot and R.R. Volkas, Phys. Rev. {\bf D 55} (1997) 5147.
\bibitem{Foot4} R. Foot and R.R. Volkas, Phys. Rev. {\bf D 56} (1997) 6653.
\bibitem{nnulimit} N. Hata et al, Phys. Rev. {\bf D 55} (1997) 540.
\end{thebibliography}
\end{document}